\documentclass[%
reprint,longbibliography,preprintnumbers,
nofootinbib,
 amsmath,amssymb,
 aps,
prb,
]{revtex4-2}
\pdfoutput=1
\usepackage[utf8]{inputenc}
\usepackage{flushend}
\usepackage{dcolumn}% Align table columns on decimal point
\usepackage{bm}% bold math
\usepackage{balance}
\usepackage{amssymb}
\usepackage{appendix}
\usepackage[normalem]{ulem}

\usepackage[colorlinks = true,
            linkcolor = blue,
            urlcolor  = blue,
            citecolor =green,
            anchorcolor = blue]{hyperref}
\usepackage{verbatim}
\usepackage{color,ulem}
\usepackage[english]{babel}

\usepackage[utf8]{inputenc}
\input Starburst.fd
\newcommand*\initfamily{\usefont{U}{Starburst}{xl}{n}}\initfamily

\newcommand{\beq}{\begin{eqnarray}}
\newcommand{\eeq}{\end{eqnarray}}
\usepackage{amsmath}
\usepackage{tikz}
\usetikzlibrary{decorations.pathmorphing}
\usetikzlibrary{shapes.misc}
\tikzset{cross/.style={cross out, draw=black, minimum size=8*(#1-\pgflinewidth), inner sep=0pt, outer sep=0pt},
cross/.default={1pt}}
\usetikzlibrary{patterns,math}
\begin{document}
\title{Eliashberg theory prediction of critical currents in superconducting thin films under DC electric fields
}
\author{Giovanni Alberto Ummarino$^{1,2}$, Alessio Zaccone$^{3,4}$, Alessandro Braggio$^{5}$, and Francesco Giazotto$^{5}$}
\address{$^1$ Dipartimento di Scienza Applicata e Tecnologia, Politecnico di
Torino, Corso Duca degli Abruzzi 24, 10129 Torino, Italy.}
\address{$^2$ National Research Nuclear University MEPhI (Moscow Engineering Physics Institute),
Kashira Hwy 31, Moskva 115409, Russia.}
\address{$^3$ Department of Physics ``A. Pontremoli'', University of Milan, via Celoria 16,
20133 Milan, Italy.}
\address{$^4$  Institute for Theoretical Physics, University of Göttingen, Friedrich-Hund-Platz 1, 37077 Göttingen, Germany.}
\address{$^5$ NEST, Istituto Nanoscienze-CNR and Scuola Normale Superiore, 56127 Pisa, Italy.}
\email{giovanni.ummarino@polito.it}

\begin{abstract}
Superconducting thin metallic films, functioning as supercurrent gate-tunable transistors, have considerable potential for future quantum electronic devices. Despite extensive research, a comprehensive microscopic quantitative mechanism that elucidates the control or suppression of supercurrents in thin films remains elusive. Focusing on NbN, a prototypical material, and starting from a phenomenological ansatz that links the critical electric field with the kinetic energy parameter
needed to break Cooper pairs, we provide a quantitative analysis of the critical current using Eliashberg theory in the dirty limit without adjustable parameters. The critical kinetic energy value is identified, corresponding to the maximum supercurrent that can flow in the thin film. The peak in supercurrent density as a function of the Cooper pairs' kinetic energy arises from the interplay between the increase in supercurrent due to increased kinetic energy and the depairing effect when the kinetic energy becomes sufficiently large. The critical value of the pair's kinetic energy is subsequently employed to estimate the critical value of an external electric field required to suppress superconductivity in the sample. This estimation is in parameter-free agreement with the experimental observations. Although the disorder reduces the temperature dependence of the gating effect on the critical current, at the same time, it increases the unscreened critical electric field needed to suppress superconductivity. This enables the proposal of methods to control and reduce the critical field value necessary to suppress superconductivity in superconducting electronics.

\end{abstract}
%\pacs{74.70.Xa, 74.20.Fg, 74.25.Kc, 74.20.Mn}
%\keywords{Field effect, thin films, Eliashberg equations}
\maketitle
%%%%%%%%%%%%%%%%%%%%%%%%%%%%%%%%%%%%%%%%%%%%%%%%%%%%%%%%%%%%%%%%%%%%%%%%%%%%%%%%%%%%%%%%%%%
\section{Introduction}
In metallic substances, the effect of an electric field is usually considered negligible because of the high density of charge carriers and the strong screening effects. Nevertheless, recent experimental findings suggest that, in superconducting materials, a sufficiently strong electric field applied in proximity to a superconducting bridge could gradually reduce the critical current of the superconducting channels. This suppression of superconductivity induced by an electric field has been observed in a variety of superconducting materials \cite{review}, primarily metallic ones, such as titanium transistor-like devices \cite{ti0,ti1,ti2} and aluminum nanodevices \cite{al1,al2,al3}. In particular, this effect has also been reported without direct current flow from the gates to the superconducting channels \cite{ti2,ionicgating}. Furthermore, the influence of electric fields on superconductivity has been a subject of scientific inquiry since $1960s$ \cite{glover}. However, the theoretical justification for specific observations \cite{fe1} and the prediction of additional observable phenomena have been formulated only in recent years \cite{fe1,mercaldo2020,fe2,fe3,fe4,fe5,subrata}.

The critical current density $J_c$, in conjunction with the critical temperature $T_{c}$ and the upper critical field $B_{c2}$, is an intrinsic critical parameter that fundamentally confines the endurance of superconductivity. Although $T_{c}$ and $B_{c2}$ are relatively straightforward to measure, $J_c$ is rarely assessed due to the technical challenges of heating the sample at high currents. The depairing or pair-breaking current has been derived within the Ginzburg-Landau theoretical framework, illustrating a temperature-dependent variation of the critical current density $J_c(T)$, particularly near the critical temperature $T_c$, expressed as $J_c(T)=J_0(1-T/T_c)^{3/2}$, where $J_0$ signifies an impurity-dependent coefficient \cite{fomin}.

This research concentrates on the critical current in thin films of thickness $L$, as comprehending the bulk critical current density requires understanding the flux lattice pinning and intricate material considerations. Conversely, this geometric configuration is more relevant for addressing matters related to the foundational superconducting state and can be used in experimental critical current experiments.
Consequently, the conclusions are more general. We determine the critical current density $J_{c}(T)$ for a thin film of thickness $L$ that is comparable to the coherence length $\xi_0$ and the magnetic penetration depth $\lambda_M$, i.e., $L \leq \xi_0<\lambda_M$. We assume that the superconductor is in the dirty limit ($\ell\ll\xi_0$) where $\ell$ is the mean free path, well within the framework of Eliashberg theory \cite{revcarbi,revmia} of strong-coupling superconductivity, which is well adapted to describe NbN superconducting thin films.

We will establish a connection between $J_{c}(T)$ and the screened critical value $E_{cr}(T)$ required to suppress the superconductivity. The connection with the externally applied electric field $E_{cr,ex}(T)$  needed to suppress the superconductivity in the thin film can be obtained following the discussion of Ref.~\emph{\cite{fomin}}.
Eliashberg's theory has adeptly accounted for the deviations in superconducting properties of actual materials from the universal predictions of BCS theory. This theoretical framework is dependent on two primary variables: the electron-phonon spectral function $\alpha^{2}F(\Omega)$, which can be measured through tunneling experiments or calculated from the first principles, and the Coulomb pseudopotential $\mu^{*}(\omega_c)$, whereby $\omega_c$ serves as a cut-off energy ($\omega_c \geq 3 \Omega_{max}$, $\Omega_{max}$ denotes the maximum phonon energy). The Coulomb pseudopotential $\mu^{*}(\omega_c)$ can be obtained indirectly through tunneling experiments or, albeit with difficulty, calculated from first principles; however, it is generally treated as a free parameter, approximately spanning between the limits described in \cite{revcarbi,revmia} and $0.1-0.3$. Once these two parameters are established, all other superconducting properties can be computed, and for conventional superconductors, all computed physical observables exhibit perfect concordance with experimental results \cite{revcarbi,revmia}. Using this approach, we can compute the critical current as a temperature- and impurity-content (disorder) function with a parameter-free theory.

In a previous paper \cite{Alessio}, we examined the same problem from a different angle: i.e., the suppression of superconductivity by an external electrostatic field. That theoretical model \cite{Alessio} treats the Cooper pair as an s-wave bound state, which is subjected to a small but non-vanishing electric field at a surface layer of thickness $\lambda_E$ quite comparable with the same metallic film layer.
Using the Schr{\"o}dinger equation solution for s-wave bound state dissociation through electric-field-assisted tunneling, an expression for the critical electric field magnitude inside the thin film layer is calculated to break the Cooper pairs, thus suppressing superconductivity.
This screened critical electric field $E_{cr}$ is proportional to the ratio between the value of the superconducting energy gap $\Delta(T)$, calculated with Eliashberg theory, and the effective coherence length $\xi(T)$ which is associated with the average size of the Cooper pair that lives in a strong inhomogeneous electric field on the metal surface. In this paper \cite{Alessio}, we also calculated that this critical electric field $E_{cr}$ is connected to the external (non-screened) critical electric field $E_{cr,ex}$.
We used an empirical input from the experiments of Piatti et al. \cite{Piatti}, for the same example of NbN thin films discussed above. According to that reference, the penetration depth of the electric field into NbN superconducting thin films is significantly greater than what is predicted by Thomas-Fermi estimates and other standard screening theories of (normal) metals, which typically
measure just a few angstroms in the range from 0.5 to 3 nm.

The purpose of this paper is to establish a phenomenological relationship between the critical supercurrent density $J_{c}$ that the material can host and the critical value of an screened electric field $E_{cr}(T)$ that must be applied to suppress superconductivity in the same sample. Although the two phenomena seem distinct, we employ the same theoretical framework: first, (i) compute the supercurrent density and its critical value, i.e., the maximum supercurrent the thin film can host. This occurs at the Cooper pair's kinetic energy critical value, $s_c$. The latter is then used (ii) to estimate the critical value of an applied electric field, $E_{cr}$, perceived by Cooper couples, to suppress superconductivity by completely field-induced depairing. This field $E_{cr}$ is connected, in a simple way \cite{Alessio}, to the external applied electric field $E_{cr,ex}$.
The underlying assumption, driven by recent work, is that the depairing effects of the supercurrent on the Cooper pairs are comparable to those of an external electric field. These two equivalent situations are schematically represented as panel (a) and panel (b) in Fig. \ref{Figure0a}, respectively.
The Eliashberg calculations are free of adjustable parameters, as the missing parameters in the theory are calibrated by matching the resistivity calculation in the normal state with experimental data from the literature.\\

\section{The model}
We will describe the physical phenomena using the simplest possible model, which, despite the approximations, still manages to grasp the fundamental physical aspects. By removing some approximations but losing simplicity, further improving the agreement with the experimental data will be possible.
\color{red}
In the standard Eliashberg theory the starting point is the definition of the Green's function
\begin{align}
G(i\omega_{n},\textbf{p})=&[i\omega_{n}Z(i\omega_{n})+(p^{2}/2m-\chi(i\omega_{n}))\rho_3 \nonumber\\
&+\Delta(i\omega_{n})Z(i\omega_{n})\rho_1\sigma_2]^{-1}
\end{align}
Here the Green's function is the Nambu-Gor'kov Green's function in $4 \times 4$
matrix notation where the $\rho_i$ and $\sigma_i$ are the Pauli matrices
operating on the electron-hole and spin spaces, respectively,
and $\rho_1\sigma_2$ is a tensor product \cite{allenth,carbiIc1} while the $\omega_{n}$ are the Matsubara energies. The components of self energy in the Green functions are $\Delta(i\omega_{n})$, i.e. the superconducting order parameter, $Z(i\omega_{n})$, i.e. the renormalization function and $\chi(i\omega_{n})$, the chemical potential shift.
After using self-consistency relations and averaging over a spherical Fermi surface, the standard Eliashberg equations for $\Delta(i\omega_{n})$ and $Z(i\omega_{n})$ are obtained \cite{allenth,revmia,parks}. The function $\chi(i\omega_{n})$ is taken to be zero \cite{revmia,parks}.
Now we use a generalization \cite{carbiIc1} of this theory for the case of a thin film supporting a supercurrent.
When a supercurrent is present, the Cooper pairs are given a finite center-of-mass momentum $q_{s}$. This $q_{s}$ leads to a boost in the quasi-particle energy. The superfluid velocity $v_{s}=q_{s}/2m_{e}$ is taken as uniform in our case, where we discuss currents in thin films. In this case, we have to generalize the Green's function in this way \cite{carbiIc1,carbiIc2,makiIc1,makiIc2,fulde}
\begin{align}
G(i\omega_{n},\mathbf{p})=&[i\omega_{n}Z(i\omega_{n})-\mathbf{v}\cdot \mathbf{q_s}+(p^{2}/2m-\chi(i\omega_{n}))\rho_3 \nonumber\\
&+\Delta(i\omega_{n})Z(i\omega_{n})\rho_1\sigma_2]^{-1}.
\end{align}
The supercurrent density $J_s(T)$ is given in terms of the finite temperature Green’s functions as \cite{carbiIc1,carbiIc2,makiIc1,makiIc2,fulde}
\begin{align}
\mathbf{J_s}(T)=\frac{ek_B T}{m_e}\sum_{\omega_{n}}\int\frac{dp^{3}}{(2\pi)^{3}}Tr(\mathbf{p} G(i\omega_{n},\mathbf{p}))
\end{align}
\color{black}
In conventional one-band s-wave superconductors, the energy gap $\Delta(T)$ remains almost unaffected by supercurrent flow up to a critical value of the superfluid velocity; at this point, pair breaking becomes possible. Beyond a critical value of $v_{s}$, the rapid reduction in the gap leads to $J_{s}=0$ for $\Delta=0$.
Our model is composed of two parts: the Eliashberg equations \cite{revcarbi,revmia}, modified to describe a thin film traveled by a supercurrent, and the expression of the supercurrent, obtained by the Green function formalism, which is solved with the solutions of the Eliashberg equations as input. These new Eliashberg equations always consist of two coupled equations: the first for the gap $\Delta(i\omega_{n})$ and the second for the renormalization functions $Z(i\omega_{n})$ while the chemical potential shift $\chi(i\omega_{n})$ is zero. In the imaginary-axis formalism, the equations \cite{carbiIc1}, when the Migdal theorem works \cite{Umma5}, read:
\begin{align}
\omega_{n}Z(i\omega_{n})&=\omega_{n}+ \pi k_{B} T\sum_{m}\Lambda(i\omega_{n},i\omega_{m})N^{Z}(i\omega_{m})+\\
&[\Gamma^{N}+\Gamma^{M}]N^{Z}(i\omega_{n})\nonumber
\end{align}
\begin{eqnarray}
&&Z(i\omega_{n})\Delta(i\omega_{n})=\pi k_{B}
T\sum_{m}\big[\Lambda(i\omega_{n},i\omega_{m})-\mu^{*}(\omega_{c})\times\nonumber\\
&&\times\Theta(\omega_{c}-|\omega_{m}|)\big]N^{\Delta}(i\omega_{m})
+[\Gamma^{N}-\Gamma^{M}]N^{\Delta}(i\omega_{n})\phantom{aaaaaa}
\end{eqnarray}
The parameters $\Gamma^{N}$ and $\Gamma^{M}$ are the scattering rates of nonmagnetic and magnetic impurities, respectively, while $\Theta$ is the Heaviside function.
In these equations, we have defined
\begin{eqnarray}
\Lambda(i\omega_{n},i\omega_{m})=2
\int_{0}^{+\infty}d\Omega \frac{\Omega
\alpha^{2}F(\Omega)}{[(\omega_{n}-\omega_{m})^{2}+\Omega^{2}]}
\end{eqnarray}
and
%
%\begin{widetext}
\begin{eqnarray}
N^{\Delta}(i\omega_{n})=\!\!\!\!\int_{-1}^{1}\!\!\!\!\!\!dz\frac{\Delta(i\omega_{n})Z(i\omega_{n})}
{2\sqrt{(\omega_{n}Z(i\omega_{n})-isz)^{2}+\Delta^{2}(i\omega_{n})Z^{2}(i\omega_{n})}}
\nonumber
\end{eqnarray}
\begin{eqnarray}
N^{Z}(i\omega_{n})=\!\!\!\!\int_{-1}^{1}\!\!\!\!\!\!dz\frac{i(\omega_{n}Z(i\omega_{n})-isz)}
{2\sqrt{(\omega_{n}Z(i\omega_{n})-isz)^{2}+\Delta^{2}(i\omega_{n})Z^{2}(i\omega_{n})}}
\nonumber
\end{eqnarray}
%\end{widetext}
%
\color{red}
The new Eliashberg equations in the exsternal structure are exactly the same of the standard theory but the definition of $N^{Z}(i\omega_{n})$ and $N^{\Delta}(i\omega_{n})$ changes lightly. In the limit of $q_s\rightarrow 0$ we refind the standard Eliashberg equations.
\color{black}
 The parameter $s=v_{F}q_{s}$ is the kinetic energy that a Cooper pair acquires when there is an electric supercurrent, and $v_F$ is the Fermi velocity.
The electron-phonon coupling constant is defined as $\lambda=2\int_{0}^{+\infty}d\Omega\frac{\alpha^{2}F(\Omega)}{\Omega}$.
In a more general situation, the solution of Eliashberg equations requires some input parameters, such as the electron-phonon spectral function $\alpha^{2}F(\Omega)$ and the Coulomb pseudopotential $\mu^{*}(\omega_{c})$. There are also non-magnetic $\Gamma^{N}$ and magnetic $\Gamma^{M}$ impurity scattering rates if disorder and magnetic impurities are present.

Under the application of a small momentum $q_s$ to the superconducting condensate in a zero magnetic field, there is no appreciable change in the BCS ground-state wave function and the superconducting condensate will be shifted from a condensate of pairs of momentum $(k,\uparrow)$ and $(-k,\downarrow)$ to one of pairs of $(k+q_s,\uparrow)$ and $(-k+q_s,\downarrow)$. The quasiparticle excitation energy will change to $s=v_F q_s$.
Following the usual development of strong coupling theory and by using the formalism of Green's function \cite{carbiIc1,carbiIc2,makiIc1,makiIc2,fulde} it is possible to obtain the equation for the superconductive density of current $J_s(T)$:
\begin{align}
&\mathbf{J_{s}}(T)=\frac{3en_0}{m_{e}v_{F}}k_{B}T \times \\
&\times \sum_{n=-\infty}^{+\infty}\!\int_{-1}^{1}\!\!\!\!\!dz\frac{i(\omega_{n}Z(i\omega_{n})-isz)z}
{\sqrt{(\omega_{n}Z(i\omega_{n})-isz)^{2}+\Delta^{2}(i\omega_{n})Z^{2}(i\omega_{n})}}\frac{\mathbf{q_{s}}}{q_{s}}\nonumber
\end{align}
where $n_0$ is the density of the carriers and $m$ is the electron mass. This equation is the expression for the supercurrent density induced in a thin film with the application of a momentum $q_s$ to the condensate.
We solve the Eliashberg equations and the expression of the supercurrent density self-consistently and find a value of $q_s=q_{sc}$ that maximizes the superconducting current density.
This value $q_{sc}$ identifies a critical value of the kinetic energy parameter $s_c$, and the critical current density is defined, in the usual way, $J_s(T,s=s_c)=J_c(T)$. Of course, the value of the kinetic energy $s_c(T)$ is similar to the energy gap $\Delta(T)$.
According to our previous paper \cite{Alessio} (see also \cite{fomin2}), the critical electric field $E_{cr}$ is proportional to the ratio between the superconducting energy gap value $\Delta$ and the effective coherence length $\xi$.
In this model, we substitute the gap value with the critical value of the kinetic energy parameter $s$ obtained from the supercurrent calculation.
The fundamental difference respect to previous modes is now we can calculate also the temperature behaviour of critical electric field and critical current density and establish criteria to be in the optimal condition (films with little disorder).
The corresponding electric field $E_{cr}(T)$ needed to split a Cooper pair \cite{Alessio}, that operates over a distance comparable to the coherence length $\xi_0(T)$ is given by \cite{Alessio,fomin2}
\begin{equation}
E_{cr}(T)=\frac{2s_c(T)}{e\,\xi_0(T)},
\end{equation}
where $s_c$ is the kinetic energy that a Cooper pair acquires when an electric field is applied.
$\xi_0(T)$ is a measure of the spatial distance between two electrons forming a Cooper pair.
However, if we study the behavior of thin films, we must remember that they are disordered materials, so we have to substitute $\xi_0(T)$ with $\xi(T)$. Furthermore, the latter is given by \cite{Alessio,fomin2}:
\begin{equation}
    \frac{1}{\xi(T)}=\frac{1}{\xi_0(T)}+\frac{1}{\ell}
\end{equation}
where $\ell$ is the mean free path.

\color{red}
\section{Equivalence between Eelectric field-induced and supercurrent-induced depairing}
\begin{figure}
\begin{center}
\includegraphics[keepaspectratio, width=0.9\columnwidth]{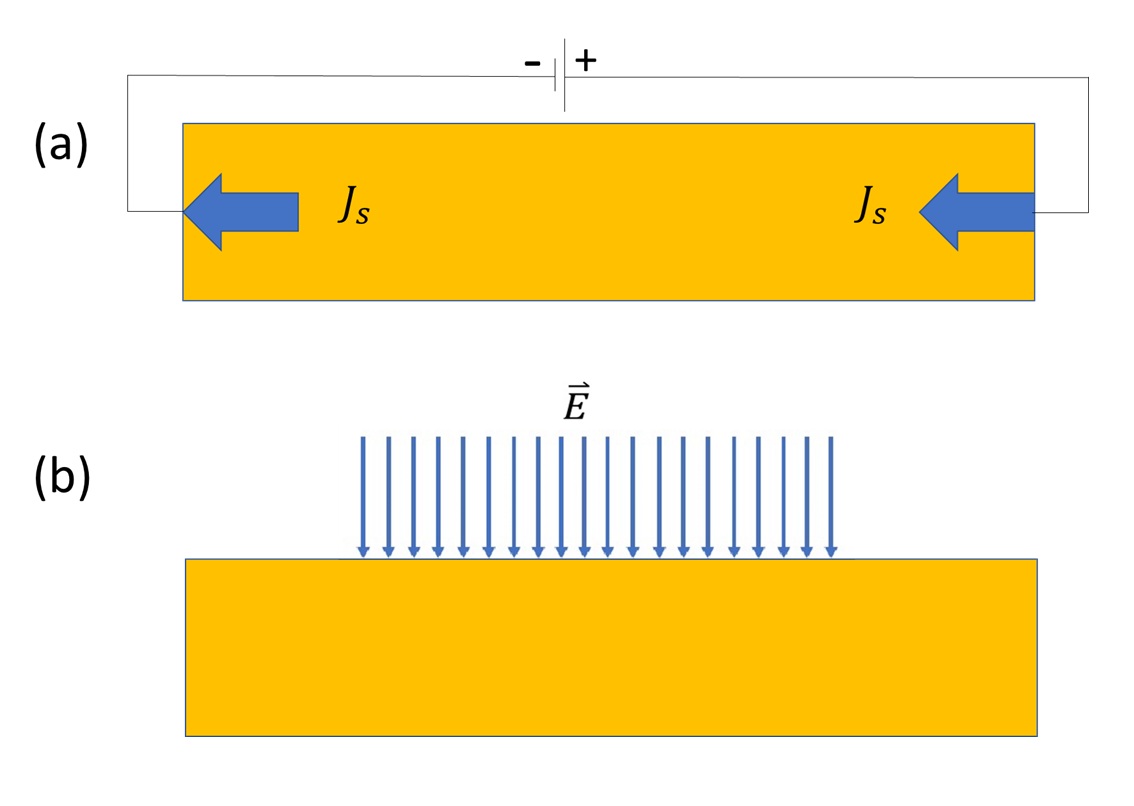}
\vspace{-5mm} \caption{(Color online)
Schematic of the two equivalent physical systems, or protocols, studied in this paper. (a) The supercurrent travels inside the thin film. (b) An external electric field applied transverse to the thin film produces the same Cooper depairing effect as the supercurrent in panel (a).
 }\label{Figure0a}
\end{center}
\end{figure}

As an ansatz, we assume that the pair-breaking effect of the supercurrent density (with no external electric field) and that of an externally applied electric field are equivalent, see Fig. \ref{Figure0a}. 

This is justified based on the following arguments. Cooper pairs in a superconductor can be broken either by increasing the kinetic energy via a supercurrent or by applying an external electric field. Both mechanisms inject energy into the condensate, and when this energy exceeds the superconducting gap $\Delta$, pairing is no longer energetically favorable.

\subsection{1. Pair breaking from supercurrent}

In the presence of a supercurrent, Cooper pairs acquire a finite center-of-mass momentum \( \mathbf{q} \), corresponding to a superfluid velocity
\[
\mathbf{v}_s = \frac{\hbar \mathbf{q}}{2m},
\]
where \( m \) is the electron mass.

The kinetic energy per Cooper pair is then
\[
s \sim E_{\text{kin}} = \frac{1}{2} M v_s^2 = \frac{\hbar^2 q^2}{4m},
\]
where \( M = 2m \) is the mass of a Cooper pair.

Pair breaking occurs when this kinetic energy becomes comparable to the superconducting gap:
\[
s \sim E_{\text{kin}} \gtrsim \Delta \quad \Rightarrow \quad \frac{\hbar^2 q^2}{4m} \gtrsim \Delta.
\]

Solving for the critical momentum and velocity we get the following estimate for the critical momentum:
\[
q_c \sim \sqrt{\frac{4m\Delta}{\hbar^2}}, \quad v_{s,c} \sim \sqrt{\frac{\Delta}{m}}.
\]

\subsection{2. Pair breaking from electric field}

An external electric field \( \mathbf{E} \) does work on the charge carriers. Over a coherence length \( \xi \) (the typical size of a Cooper pair), the energy gained by an electron is:
\[
\Delta E = e E \xi.
\]

Pair breaking occurs when this energy exceeds the gap $\Delta$:
\[
e E \xi \gtrsim \Delta \quad \Rightarrow \quad E_c \sim \frac{\Delta}{e \xi}.
\]

These estimates for the critical conditions are summarized in Table 1.

\begin{table}[h!]
\caption{Pair-breaking conditions in superconductors under different external perturbations. A supercurrent increases the Cooper pair kinetic energy, while an electric field injects energy across the coherence length. In both cases, when the energy exceeds the superconducting gap \( \Delta \), Cooper pairs are broken.}
\begin{ruledtabular}
\begin{tabular}{ll}
Mechanism & Pair-breaking condition \\ 
\hline
Supercurrent & \( \frac{\hbar^2 q^2}{4m} \gtrsim \Delta \quad \Rightarrow \quad v_s \gtrsim \sqrt{\Delta/m} \) \\
Electric field & \( e E \xi \gtrsim \Delta \quad \Rightarrow \quad E \gtrsim \Delta / (e \xi) \) \\
\end{tabular}
\end{ruledtabular}
\label{tab:pair_breaking}
\end{table}

\color{black}

Therefore, we use the critical value of the parameter $s$ to calculate the pair-breaking effect carried out by a screened electrostatic field.
In order for superconductivity to disappear, the Cooper pairs must be destroyed. To do this, the Cooper pairs must acquire a certain kinetic energy that is the same whether it comes from the electric field or from the supercurrent
When we calculate the critical current density $J_c(T)$, we find the corresponding value of the kinetic energy parameter: $s(T)=s_c(T)$ at $J_s(T)=J_c(T)$. Hence, we use the critical value $s_c$ to calculate the critical field $E_{cr}$ for depairing.
 We will discuss in which way compute the actual value (external, non screened) of the electric field that must be supplied externally to suppress the superconductivity and we refer the reader to the corresponding calculation reported in Ref. \cite{Alessio}.

%%%%%%%%%%%%%%%%%%%%%%%%%%%%%%%%%%%%%%%%%%%%%%%
%%%%%%%%%%%%%%%%%%%%%%%%%%%%%%%%%%%%%%%%%%%%%%%%%

\section{Input parameters from experimental system}
If we wish to reproduce experimental data, we must know many input parameters. For example, we consider a thin film ($L=4.5$) nm of NbN with $T_c=13.1$ $K$ ~\cite{pratap1}.
The input data for solving the Eliashberg equations are $\alpha^{2}F(\Omega)$ and $\mu^{*}(\omega_{c})$, which can be obtained by inversion of tunnel experiments \cite{revcarbi} or by density functional theory (DFT) calculations \cite{sanna}.
The electron-phonon spectral function, see inset of figure 3, $\alpha^{2}F(\Omega)$ is present in the literature \cite{a2FNbN} with a value of electron-phonon coupling costant $\lambda=1.46$ and we fix the value of the Coulomb pseudopotential $\mu^{*}(\omega_{c})=0.2866$ with $\omega_c=180$ $meV$. These two quantities $\alpha^{2}F(\Omega)$ and $\mu^{*}(\omega_{c})$, inserted in the Eliashberg equations, reproduce exactly the experimental $T_c$.
Magnetic impurities are assumed to be absent, so we put $\Gamma^{M}=0$.
The scattering rate of nonmagnetic impurities (disorder) $\Gamma^{N}$ must be determined.
$\Gamma^{N}$ can be obtained from the extrapolated resistivity value at zero temperature and from $\hbar\omega_p$ as explained in Appendix A.
From Allen's theory~\cite{Allen,Grimvall}, which corresponds to Eliashberg's theory for the normal state, we can see that $\Gamma^{N}$ is connected to the resistivity value at $T=0$ K:
$\Gamma^{N}=\varepsilon_0 \rho(T=0)\hbar \omega^{2}_p=1.24$ eV (usually $\rho(T=0)\simeq\rho_n=91 \mu\Omega cm$ \cite{pratap1}) knowing that \cite{pratap1} $\hbar\omega_p=10$ $eV$.
In this way, our calculations of the critical current do not have free parameters.
%%%%%%%%%%%%%%%%%%%%%%%%%%%%%%%%%%%%%%%%%%%%%%%%%%%%%%%%%%%%%%%%%%%%%%

\begin{figure}
\begin{center}
\includegraphics[keepaspectratio, width=0.9\columnwidth]{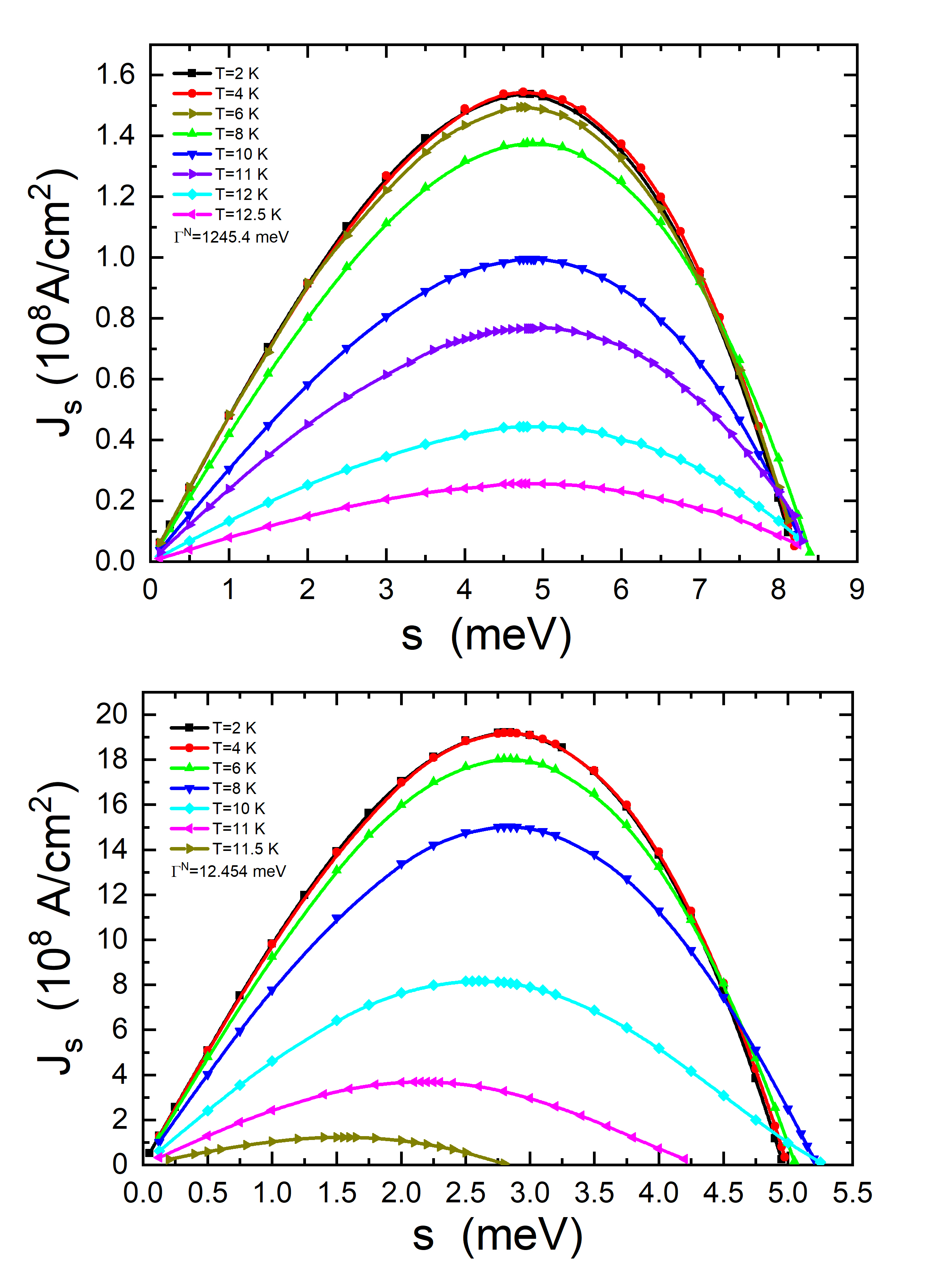}
\vspace{-5mm} \caption{(Color online)
The calculated density of supercurrent $J_s$ as a function of the parameter $s$ is shown for different temperatures. The maximum of $J_s$ corresponds to $J_c$ and identifies the critical value $s_c$~\cite{carbiIc1}. The "real case" (significant disorder) is shown in the upper panel, the "ideal case" (low disorder) is shown in the lower panel.
 }\label{Figure1a}
\end{center}
\end{figure}

%%%%%%%%%%%%%%%%%%%%%%%%%%%%%%%%%%%%%%%%%%%%%%%%%%%%%%
\section{Results}

The results of the previous model will now be discussed in detail and compared with the experimental observations.
In Fig. 2 (top and bottom panels), we first show how we calculate the critical current. First, we solve the Eliashberg equations (Eqs. 1 and 2) for different values of $s$ and then calculate the current density $J_s$ in a self-consistent way (Eq. 6). $J_c$ is the value that corresponds to the peak of $J_s$ in Fig. 2. We study two situations by placing two different values of $\Gamma^{N}$ in the Eliashberg equation.
This parameter is proportional to the disorder in the sample. The upper panel of Fig. 2 is connected with the experimental data ($\Gamma^{N}$ is obtained by resistivity measurements). In contrast, the bottom panel refers to a hypothetical "ideal case" where the disorder, i.e., $\Gamma^{N}$, is reduced by a factor of 100.
In correspondence with the maximum density of the superconducting current $J_s$, we find a critical value of $s$~\cite{carbiIc1}.
As shown in the top panel of Fig. 2, $s_c$ is practically independent of the temperature (real case, with significant disorder). In contrast, the dependence on the temperature is evident in the bottom panel (ideal case with low disorder).
Furthermore, the values of the Cooper pair $s$ kinetic energy for the critical current increase significantly for the real case. This implies that the expected critical electric field is higher for the real disordered case than for the ideal case. This situation is probably similar to an increase in the upper critical field with disorder in low $T_c$ superconductors without strongly affecting $T_c$.
%%%%%%%%%%%%%%%%%%%%%%%%%%%%%%%%%%%%%%%%%%%%%
\begin{figure}
\begin{center}
\includegraphics[width=\columnwidth]{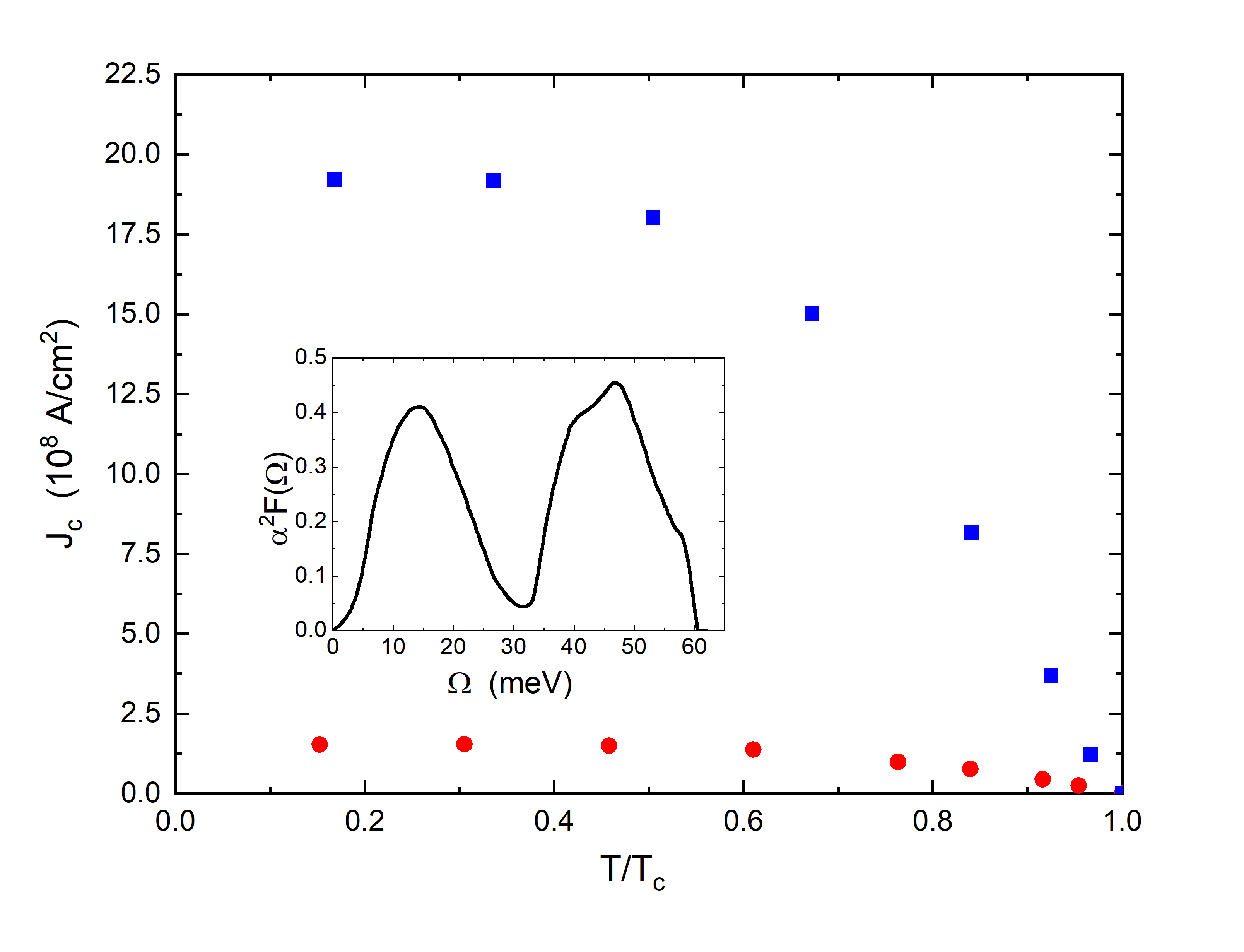}
\vspace{-5mm} \caption{(Color online)
The calculated critical density of supercurrent $J_c$ as a function of temperature in the real case (red full circles) and ideal case (dark blue squares) case. The inset shows the electron-phonon Eliashberg spectral function $\alpha^{2}F(\Omega)$ of NbN.
 }\label{Figure2a}
\end{center}
\end{figure}
%%%%%%%%%%%%%%%%%%%%%%%%%%%%%%%%%%%%%%%%%%%%%%%%%%

It is interesting to consider the behavior of the critical current density as a function of temperature for the two cases. Figure 3 shows the behavior of $J_c(T)$ as a function of temperature again for these two values of non-magnetic disorder $\Gamma^{N}$. The red full circles refer to the real disordered case,
%experimental data
while the dark blue full squares refer to the ideal clean case, where the disorder is reduced by a factor of 100. We can observe that reducing the disorder increases the value of the critical current. In this case, by a factor of 10.
%%%%%%%%%%%%%%%%%%%%%%%%%%%%%%%%%%%%%%%%%%
\begin{figure}
\begin{center}
\includegraphics[width=\columnwidth]{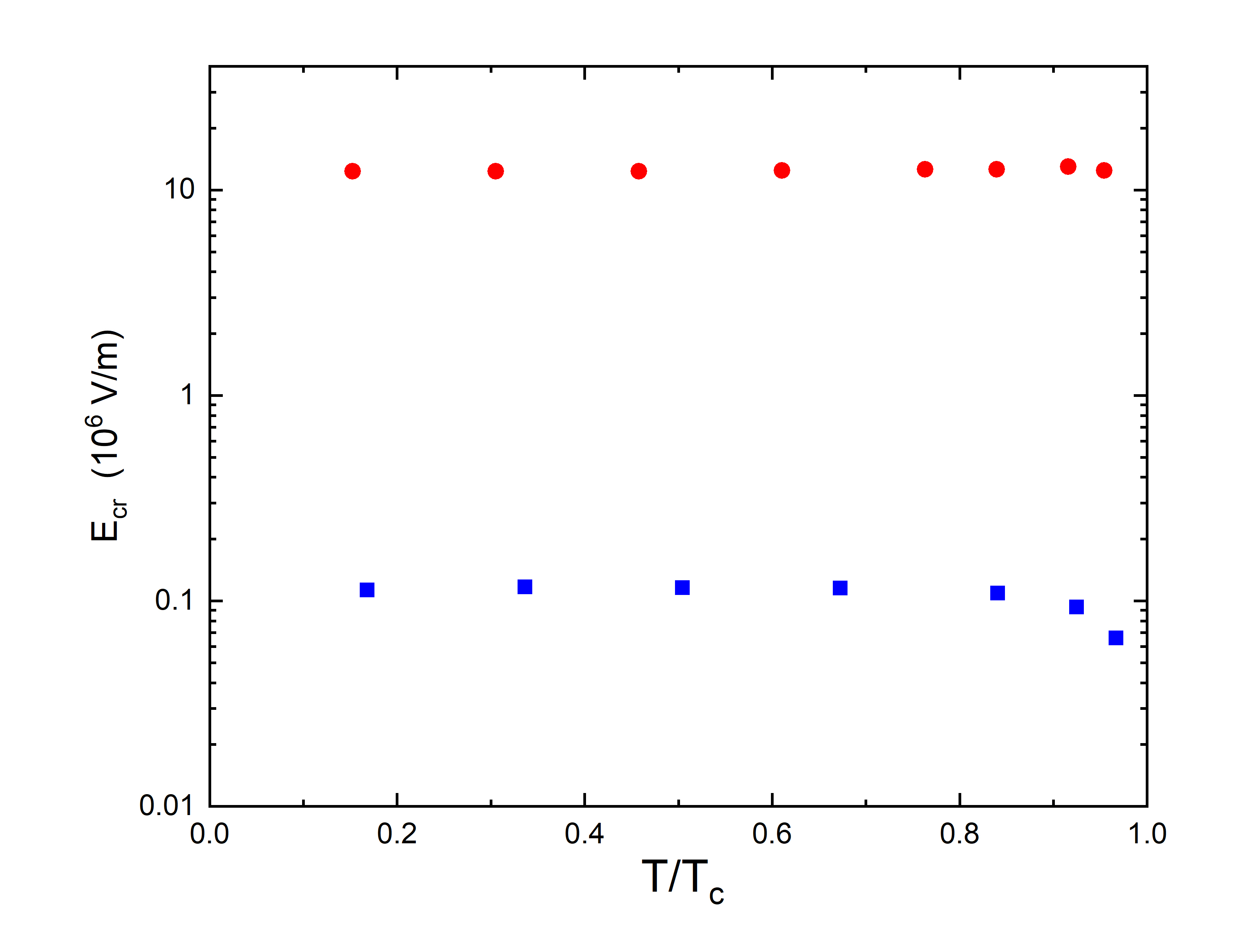}
\vspace{-5mm} \caption{(Color online)
The calculated critical field $E_{cr}(T)$ as a function of temperature in the real case, with significant disorder (red full circles) and the ideal case, with low disorder (dark blue circles).
 }\label{Figure3a}
\end{center}
%\label{fig:Ecr}
\end{figure}
%%%%%%%%%%%%%%%%%%%%%%%%%%%%%%%%%%%%%%%%%%%%%%%%%%%%%%%%%%

Subsequently, we use the critical value $s_c$ to calculate the critical field $E_{cr}$ following the phenomenological ansatz of Eq.\ref{eq:EcrDis}. This is shown in Fig.\ref{Figure3a} as a function of the temperature ratio $T/T_c$. Also, in this case, the red full circles refer to the real disordered case, while the dark blue full squares refer to an ideal case with very low disorder. Here we see that the disorder increases the value of the critical field (in this case, by a factor of 100). We can notice the different behavior of the critical field in the two cases: in the real-world case, with disorder, the critical field goes abruptly to zero near the critical temperature, whereas in the ideal case (low disorder), the decrease is smoother. So, disorder not only increases the required values of the critical field, but also proportionally reduces the temperature dependence of the critical field with respect to the temperature if it is very close to the superconducting critical temperature.

%%%%%%%%%%%%%%%%%%%%%%%%%%%%%%%%%%%%%%%%%%%%%
\begin{figure}
\begin{center}
\includegraphics[ width=0.9\columnwidth]{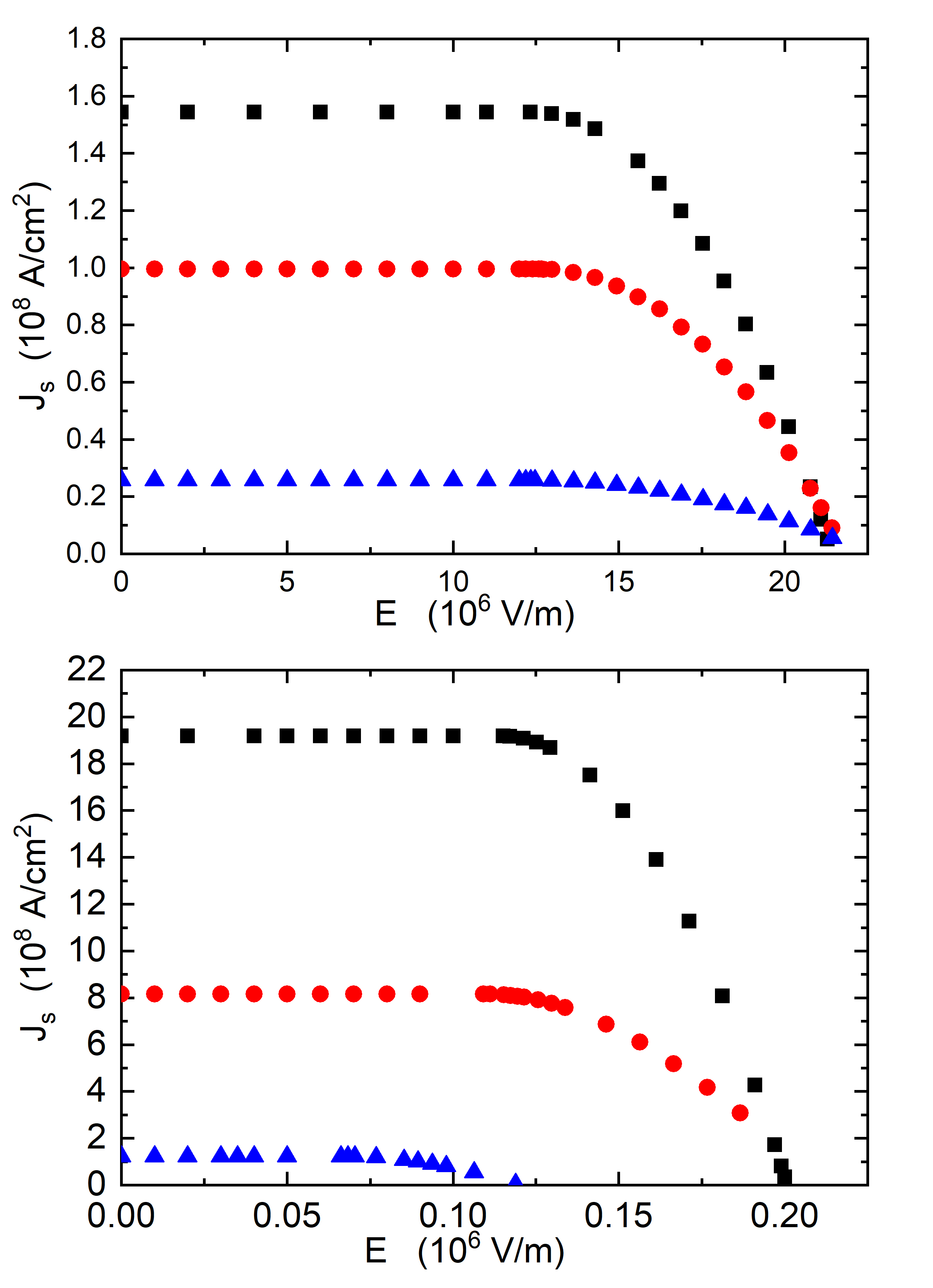}
\vspace{-5mm} \caption{(Color online)
Upper panel: The calculated density of supercurrent $J_s$ is linked to the electric field at $T=4.0$ $K$ (filled black squares), $T=10.0$ $K$ (filled red circles), and $T=12.5$ $K$ (filled dark blue triangles).
Lower panel: The calculated density of supercurrent $J_s$ is linked to the electric field at $T=4.0$ $K$ (filled black squares), $T=10.0$ $K$ (filled red circles), and $T=11.0$ $K$ (filled dark blue triangles) in the ideal case.
 }\label{Figure4a}
\end{center}
\end{figure}
%%%%%%%%%%%%%%%%%%%%%%%%%%%%%%%%%%%%%%%%%%%%%%%%%%%%%%%%%%%%%%%%%%%%%%%%%

Now, it is interesting to investigate how the electric field could affect the critical current.
We suppose that the link between these two quantities $E=|\vec{E}|$ and $s$ is schematically given by:
\begin{equation}
    s(E)=\frac{s_c}{E_{cr}}E\,\theta(E_{cr}-|E|) \label{theta}
\end{equation}

where $\theta(x)$ is the Heaviside step function.
The physical meaning of this relation is that the kinetic energy of the Cooper pairs in the condensate, due to the dissipationless current flowing, for values greater than $s_c$ would generate depairing. Equivalently, we assume that an external electric field of magnitude $E$ applied across the thin film would generate an effective kinetic energy term which will become critical at $E_{cr}$. For simplicity, we assume a linear relationship between $s$ and $E$,
where the straight line must pass through the point $s_c = E_{cr}$. Notably, we computed the effects of the module of the electric field, since the depairing mechanism is independent of the electric-field orientation (ingoing or outgoing from the metallic surface). It is related to the inherent bipolarity of the effect, substantially reported in many experiments\cite{PaolucciBipolar}.
This assumption is the simplest one can make.
%Below a threshold value $E_{cr}$, nothing happens, while after that, the dependence is linear.
Of course, in our theory, $E$ and $E_{cr}$ are of absolute value. Of course, here the electric field $E_{cr}(T)$ is the electric field that penetrates the metal and acts directly on the Cooper pairs, and not the external electric field $E_{cr,ex}(T)$ renormalized by the screening effect which will be much higher \cite{Alessio}.
\color{red}
We do a short summary of the idea of Zaccone et al \cite{Alessio}.
The starting point is the consideration that the magnitude of the electric field within the thin film is significantly lower than that of the external electric field and exhibits spatial heterogeneity because of the penetration profile. This means that the magnitude of electric field in the thin film is significantly lower than the external one due to screening. We use empirical input from ab-initio simulations and experiments of Piatti et al. \cite{Piatti}, for the same example of NbN thin films discussed above.
According to that reference, the penetration depth $\lambda_E$ of the electric field into NbN superconducting thin films is significantly greater than what is predicted by Thomas-Fermi estimates.
According to the screening theory of normal metals, see references in the paper of Zaccone et al \cite{Alessio}, the electric field incident onto the surface of a metal decays exponentially from the value it has at the interface. Denote by $E_{0}$ the value of electric field at the interface between the metal and the surrounding environment and by $x=0$ the coordinate of the interface. Hence, $E_{0}\equiv E(x=0)$. We assume, for simplicity, that $E_{0}$ coincides with the magnitude of the external electric field as determined in the experiment.
We used this semiempirical form for the profile of the electric field inside the sample \cite{Alessio}:
\begin{equation}
E(x)=E_0 \exp(- x/\lambda_E),
\end{equation}
 For NbN in the superconducting state, this length is large (compared to normal metals), that is, $\lambda_E \approx 0.5-3$ nm, as shown in \cite{Piatti}.
 We assume a bipolar setup in which the electric field is incident on both sides (surfaces) of the film. This implies that the electric field is exponentially decaying within the film from its external value $E_0$, on both sides at $z=0$ and $z=L$. Hence, we have complete symmetry across the midline $z=L/2$ plane of bilateral symmetry. This simplifies our problem, as we only need to consider the exponential decay on one side of the film. We can safely assume that the superconductor sees the inhomogeneous electric field mediated on the scale of at least $\xi$. However, it is assumed that the thickness of the film is not larger than the coherence length. We can thus estimate the effective electric field as the average magnitude of the electric field inside the whole sample as
\begin{equation}
\bar{E}=\frac{1}{L/2}\int_{0}^{L/2}E_0\, e^{-x/\lambda_E} dx = E_0 \frac{2\lambda_E}{L}(1- e^{-L/(2\lambda_E)}). \label{decay}
\end{equation}
In this formula, $E_0$ represents the external incident electric field on the surface of the film, while $\bar{E}$ represents the average value of the electric field seen by the superconductor inside the film.
Hence, we must set $\bar{E}$ equal to $E_{cr}$  and solve for $E_0$, the value we supply externally to suppress the superconductivity. Let us call this value $E_0 \equiv E_{cr,ext}$ to distinguish it from $E_{cr}$ used before.
The magnitude of external critical electric field is thus given by
\begin{equation}
E_{cr,ex}=\frac{E_{cr}}{ \frac{2\lambda_E}{L}(1- e^{-L/2\lambda_E})}. 
\end{equation}
This means that, if $\lambda_E=1$ nm \cite{Alessio} and $L=4.5$ nm we have to multiply our value by 2.5.
Thin films, such as those used in supercurrent field effect devices, have a microstructure characterized by microcrystallites, the size of which limits the value of $\ell$. Since, typically, $\ell \ll \xi_0(T)$ (because $\xi_0(T)$ can be tens or hundreds of nanometers, in the case examined in this paper (NbN) $\xi_0(T)=\frac{\hbar v_F}{\pi\Delta}\sim 150$ nm and $\ell\sim0.75$ nm ~\cite{pratap1}, the coherence length $\xi$ is controlled by $\ell$ and therefore by disorder and $\xi \approx \ell$. In the case of NbN $v_F=1.55\cdot 10^{6}$ m/s and the gap $\Delta$ is calculated by solution of Eliashberg equations, $\Delta=2.3$ meV.
We chose the NbN superconductor because it is essential for applications, all input parameters are well known, and it is well within the dirty limit.
\color{black}
Since for experimental NbN systems the disorder is always present, in the form of small grains (crystallites) that are randomly packed \cite{Sidorova}, we have $\xi \approx \ell$, and, therefore \cite{Alessio}:
\begin{equation}
E_{cr}=\frac{2s_c(T)}{e\,\xi(T)}\simeq\frac{2s_c(T)}{e\,\ell}.
\label{eq:EcrDis}
\end{equation}

Figure 5 shows the link between $J_{s}$ and $E$ for different temperatures (different colors) in the real disordered case (top panel) and the ideal low-disorder case (bottom panel).
Note that even if the critical field is almost independent of temperature in the real case, temperature dependence manifests itself in the ideal case. This is another proof that disorder makes the superconducting state more robust, obviously in the case we are studying, the low $T_c$ s-wave superconductor.

We examined a real disordered case and an "ideal case" (low disorder) to define a strategy to control the gating effects of the superconducting properties of the thin films employed in superconducting electronics.
Indeed, this analysis suggests that one needs to reduce the disorder to reduce the critical electric field. This occurs when the resistivity in the normal state at $T_c$ is very low, for example $\rho_n'=\rho_n/100$.
This means $\ell'=100\ell$ ($\ell=mv_F/ne^{2}\rho_n$) and $\Gamma^{N'}=\Gamma^{N}/100$: this means that, in this case, the thin film is of excellent quality, although we are still in the dirty limit where the theory works.
This happens because the mean free path is proportional to $\rho_n^{-1}$, but the resistivity is proportional to $\Gamma^{N}$. In this way $\ell$ is proportional to $(\Gamma^{N})^{-1}$. If we reduce $\Gamma^{N}$ of a factor 100 the means free path increases of a factor 100. In the experimental situation, $\ell=0.75$ nm, while in the ideal case we will have $\ell'=75$ nm. The coherence lenght $\xi_0(T)$ is $150$ nm so we remain in the dirty limit.

In the ideal case (low-disorder), the critical field $E_{cr}$ is reduced by two orders of magnitude, while the critical current increases by more than one order of magnitude.
The recipe for reducing the critical field is straightforward: we must make thin films with as little disorder as possible.
Magnetic impurities could also be introduced into our theory. We have not examined this case because it would drastically reduce the film's critical temperature.

\section{Simultaneous presence of supercurrent and external electric field}

Finally, we consider the situation where both contributions, i.e., the supercurrent and the electric field, are present, which is a realistic case for the supercurrent field-effect transistors.
The critical value of $s$, in this case, is the sum of two contributions:
\begin{equation}
s=s_J+s_E,
\end{equation}
where $s_J$ refers to the contribution from the supercurrent boost and $s_E$ refers to the contribution from the external electric field.
In the above formulation, $s_J$ represents the quote of the kinetic energy parameter, which is associated with the supercurrent. Hence, the kinetic energy parameter activated by the external electric field cannot carry over to maintain the supercurrent and, therefore, has to be subtracted off when considering the dependence of $J_s$ on $s$, in this case.
For all these reasons, now the maximum of the current density (cf. Fig. 2) will occur at a lower value of $s$ compared to the case where the external electric field is switched off. In practice, we have that the critical supercurrent will occur for a critical value of the kinetic energy parameter given by:
\begin{equation}
s'_c=s_c(1-E/E_{cr}).
\end{equation}
We see that the critical value of $s$ is reduced with respect to the case without the external electric field, and
so is also reduced the value of critical current density, as can be seen in Fig. 2 when we are in the case $s<s_c$ (the curves in Fig. 2  "translate" along the x-axis such that $s_c$ becomes $s'_c$). Of course, when $E=E_{cr}$, the critical supercurrent density value is zero, consistent with the fact that $E_{cr}$ is the value of an external electric field that completely suppresses the superconductivity of the thin film.
%%%%%%%%%%%%%%%%%%%%%%%%%%%%%%%%%%%%%%%%%%%%
\section{Conclusions}
The evolution of the critical current density in prototypical NbN thin films was computed as a function of temperature under the dirty limit for two contrasting disorder strengths. Although the model uses a phenomenological ansatz to connect the critical supercurrent density with the critical electric field, it lacks adjustable parameters, particularly when applied to conventional superconductors, where the theoretical input data are known or can be readily derived from experimental observations.
The supercurrent density attains a maximum with respect to the Cooper pair kinetic energy parameter $s$. This peak emerges because of the balance between the increase in current with enhanced kinetic energy of the Cooper pairs and the onset of the depairing effect, which occurs when the kinetic energy of the Cooper pairs becomes sufficiently large. The maximum point in the supercurrent delineates a critical energy parameter $s_c$, which is later used to estimate the unscreened  critical value of the electric field. After computing the correction of the screening, we can compute the external electric field necessary to suppress superconductivity in the thin film. This calculated value is consistent with experimental data from the literature and confirms a different estimate recently reported in Ref. \cite{Alessio}.
The fundamental difference with previous calculations \cite{Alessio} is that with this model we can calculate the temperature trend not only of the critical field but also of the supercurrent density and establish criteria to be in the optimal condition (films with little disorder). The previous model is more useful when we have
to work with very thin film because the film thickness appears in the theory.
The analysis further indicates that, within the framework of this model, the suggested method to diminish the critical electric field required for the suppression of superconductivity in the film is a maximal reduction of disorder. Parameter-free theoretical calculations undergo rigorous quantitative validation by obtaining a significant value for the critical electric field and demonstrating the capability to represent the resistivity in the normal state accurately.

In conclusion, extending this theory to multiband superconductors \cite{carbiIc2,ummator1,ummator2,ummator4,ummator5} or to alternative symmetries \cite{makiIc2,ummator3} of the order parameter is a direct process and can be undertaken in subsequent research studies.
\appendix
\section{Resistivity as a function of temperature in the normal state}
Finally, we want to demonstrate that our calculations are quantitatively consistent using different experimental data \cite{pratap1}.
The above calculations have been performed using parameters of the theory that must be quantitatively consistent with predictions of the same theory compared with experimental observables.

In this way, we wish to calibrate the theory parameters to reproduce the temperature dependence of resistivity in the normal state for our metallic superconductor, using precisely the same input parameters of the theory presented in the previous sections. To this aim, we use Allen's theory, which is closely connected with Eliashberg's theory ~\cite{Allen,Grimvall}. The resistivity as a function of temperature in the normal state is ~\cite{LiFeAs,resumma}
\begin{equation}
	\rho(T)=\hbar\frac{\Gamma+W(T)}{\varepsilon_0(\hbar\omega_{p})^{2}}
\label{rho}
\end{equation}
In this equation, $\omega_{p}$ is the bare plasma frequency while
\begin{equation}
	W(T)=4\pi k_BT\int_0^\infty d\Omega
			\left[\frac{\Omega/2k_BT}{\sinh\big(\Omega/2k_BT\big)}\right]^2
			\frac{\alpha_{tr}^2F_{tr}(\Omega)}{\Omega}.\\
\label{W1}
\end{equation}
In the last equations, we add the impurity contributions: $\Gamma=\Gamma^{N}+\Gamma^{M}$, that is, we add the nonmagnetic (N) and magnetic (M) impurity scattering rates.
Here, $\alpha^2_{tr}(\Omega)F_{tr}(\Omega)$ is the transport electron-phonon spectral function related to the Eliashberg function~\cite{Allen}.\\
To highlight the value of the electron-phonon coupling, we can write the spectral transport function in this way:
$\alpha_{tr}^{2}F_{tr}(\Omega)=\lambda_{tr}\alpha_{tr}^{2}F^{n}_{tr}(\Omega)$
where $\alpha_{tr}^{2}F^{n}_{tr}(\Omega)$ is a normalized spectral function with an electron-phonon coupling constant $\lambda_{n}=2\int_{0}^{+\infty}\frac{\alpha^{2}F^{n}(\Omega)}{\Omega}d\Omega$ equal to $1$.
From Allen's theory, we know \cite{Allen} that the superconducting and transport spectral functions are similar.
To simplify our model, we assume that the shapes of the spectral functions (superconductive and transport) are equal, apart from the region near $\Omega=0$. Indeed, they differ just by a scaling factor, the coupling constant, and~\cite{Allen} $\lambda_{tr} < \lambda_{sup})$. The key difference manifests itself in the behavior for $\Omega\rightarrow 0$, where the transport function behaves like $\Omega^4$ instead of $\Omega^2$ as in the superconducting state.

\begin{figure}
\begin{center}
\includegraphics[width=\columnwidth]{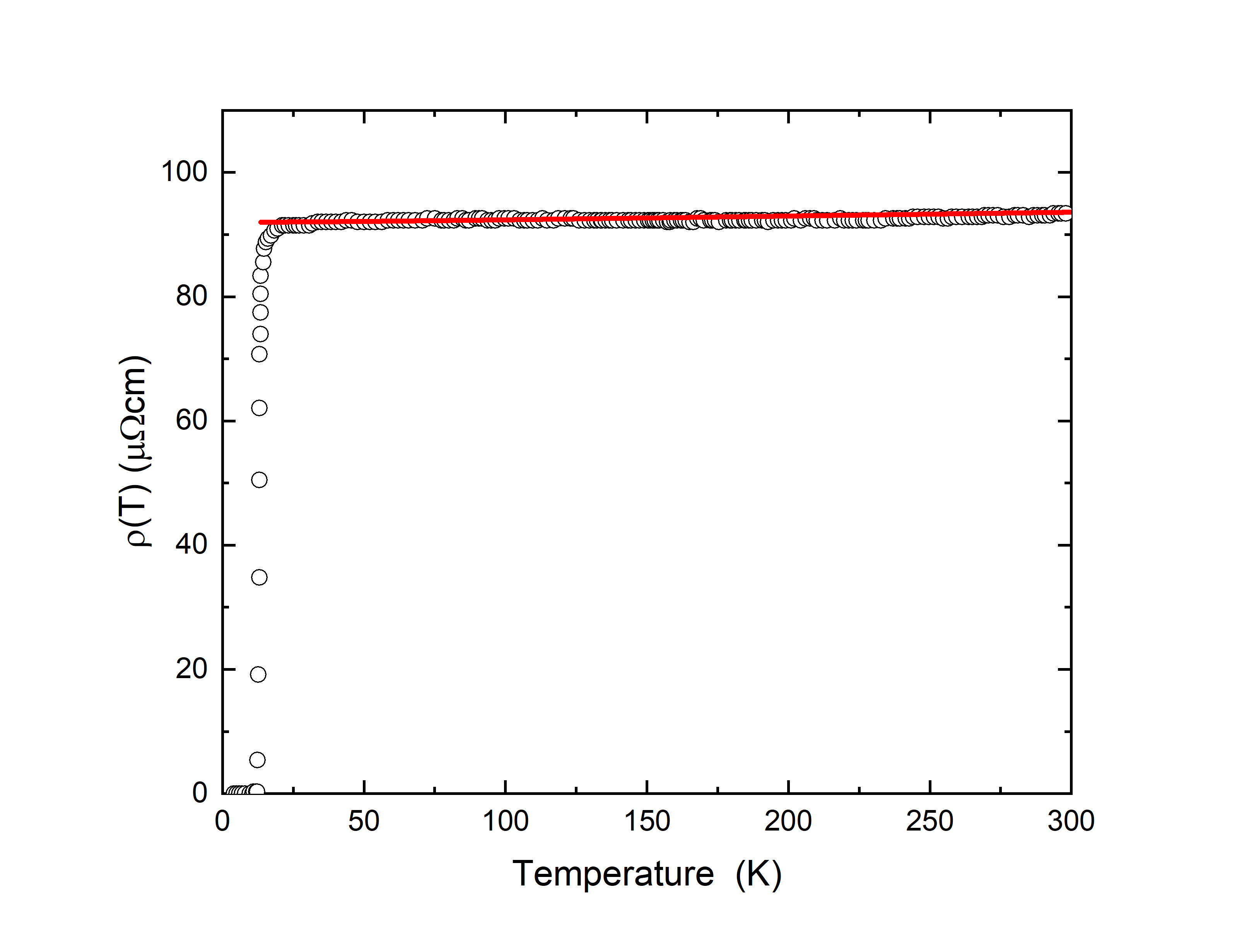}
\vspace{-5mm} \caption{(Color online)
The calculated resistivity as a function of temperature (red lines) and experimental data \cite{pratap1} (open circles).
 }\label{Figure5n}
\end{center}
\end{figure}

In low-temperature superconductors the condition $\alpha^2_{tr}(\Omega)F_{tr}(\Omega)\propto\Omega^4$ takes place
in the range ~\cite{Allen} $0 <\Omega< \Omega_{D}$, with $\Omega_{D}\approx\Omega_{ln}/10\simeq 1.5$ meV (here $\Omega_{ln}=\exp(\frac{2}{\lambda}\int_{0}^{+\infty}d\Omega \ln(\Omega)\alpha^{2}F(\Omega)/\Omega)$ being the typical phonon energy of the material). In this way, the final definition of the transport spectral function is
$\alpha^2_{tr}(\Omega)F_{tr}(\Omega)=b\Omega^{4}\vartheta(\Omega_{D}-\Omega)+c\alpha^2(\Omega)F(\Omega)\vartheta
(\Omega-\Omega_{D})$.
The constants $b$ and $c$ have been fixed by requiring continuity in $\Omega_{D}$ and the normalization condition. We chose, when $\Omega>\Omega_D$, for the $\alpha^2_{tr}(\Omega)F_{tr}(\Omega)$ the same functional form of the superconducting state. From the fit of the experimental data in Fig. 6, we find $\lambda_{tr}=0.12 <\lambda_{sup}$, as it has to be \cite{Allen}. $\lambda_{tr}$ is the only free parameter of the theory and could be calculated from first principles upon comparison with this value obtained by the resistivity fit. The result is shown in Fig. 6.
This calculation further checks the quantitative consistency of the model because the same parameter is used to produce all the predictions shown in this paper and the resistivity fit in Fig. 6.
\section{ACKNOWLEDGMENTS}
G.A.U. acknowledges partial support from the MEPhI and V. Fomin for the valuable discussions.
F.G. acknowledges the EU’s Horizon 2020 Research and
Innovation Framework Programme under Grants No. 964398 (SUPERGATE), No. 101057977 (SPECTRUM),
and the PNRR MUR project PE0000023-NQSTI for partial financial support.
A.Z. gratefully acknowledges funding from the European Union through Horizon Europe ERC Grant number: 101043968 ``Multimech'', from US Army Research Office through contract nr. W911NF-22-2-0256, and from the Nieders{\"a}chsische Akademie der Wissenschaften zu G{\"o}ttingen in the frame of the Gauss Professorship program.
A.B. acknowledges the
MUR-PRIN 2022—Grant No. 2022B9P8LN-(PE3)-Project NEThEQS “Non-equilibrium coherent thermal effects in quantum systems” in PNRR Mission 4-Component 2-Investment 1.1 “Fondo per il Programma Nazionale di Ricerca e Progetti di Rilevante Interesse Nazionale (PRIN)” funded by the European Union-Next Generation EU and CNR project QTHERMONANO.\\
%%%%%%%%%%%%%%%%%%%%%%%%
\\
%BILBLIOGRAFIA

%%%%%%%%%%%%%%%%%%%%%%%%

\end{document}